\documentclass[aps,prl,twocolumn,superscriptaddress]{revtex4-1}
\usepackage{amsmath,amsfonts,amssymb,bm}
\usepackage{graphicx}
\usepackage{xcolor}
\usepackage{soul}

\newcommand{\eq}[1]{Eq.~(\ref{#1})} 
\newcommand{\ignore}[1]{}

\begin{document}

\title{Bogoliubov Fermi Surfaces in Superconductors
with Broken Time-Reversal Symmetry}

\author{D. F. Agterberg}
\email{agterber@uwm.edu}
\affiliation{Department of Physics, University of Wisconsin,
                 Milwaukee, WI 53201, USA}
\author{P. M. R. Brydon}
\email{philip.brydon@otago.ac.nz}
\affiliation{Department of Physics, University of Otago, P.O. Box 56,
             Dunedin 9054, New Zealand}
\author{C. Timm}
\email{carsten.timm@tu-dresden.de}
\affiliation{Institute of Theoretical Physics, Technische
  Universit\"at Dresden, 01062 Dresden, Germany}

\date{\today}

\begin{abstract}
It is commonly believed that in the absence of disorder or an external
magnetic field, there are three possible types of superconducting
excitation gaps: the gap is nodeless, it has point nodes, or it has
line nodes. Here, we show that for an even-parity nodal superconducting
state which spontaneously breaks time-reversal symmetry, the low-energy
excitation spectrum generally does not belong to any of these
categories; instead it has extended Bogoliubov Fermi surfaces. These
Fermi surfaces can be visualized as two-dimensional surfaces generated
by ``inflating'' point or line nodes into spheroids or tori,
respectively. These inflated nodes are topologically protected from
being gapped by a $\mathbb{Z}_2$ invariant, which we give in terms of
a Pfaffian. We also show that superconducting states possessing these
Fermi surfaces can be energetically stable. A crucial ingredient in our
theory is that more than one band is involved in the pairing; since all
candidate materials for even-parity superconductivity with broken
time-reversal symmetry are multiband systems, we expect these
$\mathbb{Z}_2$-protected Fermi surfaces to be ubiquitous.
\end{abstract}

\maketitle

\textit{Introduction}.---The theory of superconductivity is
conventionally formulated in terms of pairing of  spin-$1/2$
fermions \cite{vol85,sig91}. The complications introduced by
additional electronic degrees of freedom, e.g., orbitals, are
not usually thought to qualitatively alter the physics.
This picture has recently been challenged for a number of
materials. For iron-based superconductors, the role
of interorbital pairing is attracting increased attention
\cite{GSZ10,NGR12,NKT16,ong16}.
Another example is the nematic superconductivity of
Cu$_x$Bi$_2$Se$_3$~\cite{CuBi2Se3}, where the odd parity of the gap is
encoded in the orbital degrees of freedom. Furthermore, theories of
pairing in YPtBi and UPt$_3$ based on $j=3/2$ and $j=5/2$ fermions,
respectively, have greatly enriched the allowed superconducting
states~\cite{bry16,NoI16}.

In this Letter, we show that the presence of  multiple bands
qualitatively changes the nodal structure of a
time-reversal-symmetry-breaking (TRSB) superconductor. Specifically,
the expected line or point nodes of an even-parity superconducting
gap \cite{vol85,sig91} are replaced by two-dimensional Fermi
surfaces of Bogoliubov quasiparticles, which are topologically
protected by a $\mathbb{Z}_2$ invariant. We further interpret these
Fermi surfaces in terms of a pseudomagnetic field arising from
interband Cooper pairs,  here referred to as
``interband pairing''.

Our conclusions are relevant for a wide range of candidate
TRSB superconductors, such as UPt$_3$~\cite{luk93,joy02,sch14,NoI16},
Th-doped UBe$_{13}$ \cite{hef90,zie04}, PrOs$_4$Sb$_{12}$
\cite{aok03}, Sr$_2$RuO$_4$ \cite{luk98,xia06}, URu$_2$Si$_2$
\cite{kas07,sch15}, SrPtAs \cite{bis13}, and Bi/Ni bilayers~\cite{gon16}. Remarkably, signatures
of these Fermi surfaces may have already been observed in Th-doped
UBe$_{13}$ \cite{zie04} (and possibly in  UPt$_3$ \cite{joy02}),
where there is evidence for  a non-zero density of states at
zero temperature, which appears not to be  due to impurities.
In addition to these known superconductors, theory has predicted TRSB
superconductivity in graphene \cite{nan12,abs14}, the half-Heusler
compound YPtBi \cite{bry16}, water-intercalated sodium cobaltate
Na$_x$CoO$_{2}\cdot y$H$_2$O \cite{sar04,kie13}, Cu-doped TiSe$_2$
\cite{gan14}, and monolayer transition metal
dichalcogenides \cite{hsu16}. A common feature of all these materials
is that the electronic structure  involves multiple bands, and so we
expect our theory to apply.

\textit{Model}.---Our starting point is a general Hamiltonian with
time-reversal and inversion symmetries containing four electronic
degrees of freedom at each momentum. These four degrees of freedom
can arise from either the combination of spin $1/2$ and two orbitals
of equal parity, or from fermions with effective angular momentum
$j=3/2$.  We work with a $j=3/2$ generalized Luttinger-Kohn
Hamiltonian \cite{lut55} in order to make our arguments most
transparent.
Although these two descriptions have different symmetry properties,
we show in the Supplemental Material (SM) \cite{suppinfo} that they
can be unitarily transformed into each other and that our model
represents a generic two-band theory including all
symmetry-al\-lowed crystal-field and spin-orbit-coupling terms. The
general form of the normal-state Hamiltonian is
\begin{align}
H_N &= c_0 1_4
  +c_{yz}\frac{J_y J_z+J_z J_y}{\sqrt{3}}
  +c_{xz}\frac{J_x J_z+ J_z J_x}{\sqrt{3}} \nonumber \\
&+c_{xy}\frac{J_x J_y+J_y J_x}{\sqrt{3}}
  +c_{3z^2-r^2}\frac{2J_z^2-J_x^2-J_y^2}{3} \nonumber \\
&+c_{x^2-y^2}\frac{J_x^2-J_y^2}{\sqrt{3}} ,
\label{Ham0}
\end{align}
where $1_4$ is the $4\times4$ unit matrix,  and the
$J_i$ are spin-$3/2$ matrices  given in the
SM \cite{suppinfo}. The coefficients $c_i=c_i({\bf k})$
of the matrices in~\eq{Ham0} are real, satisfy
$c_i({\bf k})=c_i(-{\bf k})$, and transform  in the same way
as the corresponding matrix under spatial symmetries.
$H_N$ has twofold degenerate eigenvalues
$\epsilon_\pm = c_0
  \pm (c_{yz}^2+c_{xz}^2+c_{xy}^2+c_{3z^2-r^2}^2+c_{x^2-y^2}^2)^{1/2}$.
For definiteness, we discuss the case of only one of the two bands
crossing the chemical potential so that there is only one normal-state
Fermi surface; if there are two, identical arguments pertain to both.
While our conclusions are general, numerical results are given for
the spherically symmetric Hamiltonian
$H_N = \alpha k^2 + \beta\, (\mathbf{k}\cdot\mathbf{J})^2 - \mu$
\cite{lut55}, where $\alpha$ and $\beta$ are constants ($\beta$ is
the spin-orbit coupling) and $\mu$ is the chemical potential, which
leads to $c_0=(\alpha+5\beta/4)k^2-\mu$,
$c_{3z^2-r^2}=\beta[k_z^2-(k_x^2+k_y^2)/2]$,
$c_{x^2-y^2}=\sqrt{3}\beta(k_x^2-k_y^2)/2$,
$c_{yz}=\sqrt{3}\beta k_yk_z$, $c_{xz}=\sqrt{3}\beta k_xk_z$,
and $c_{xy}=\sqrt{3}\beta k_xk_y$.

\begin{figure}[tb]
\begin{center}
\vspace{0.5cm}
\includegraphics[width=0.8\columnwidth]{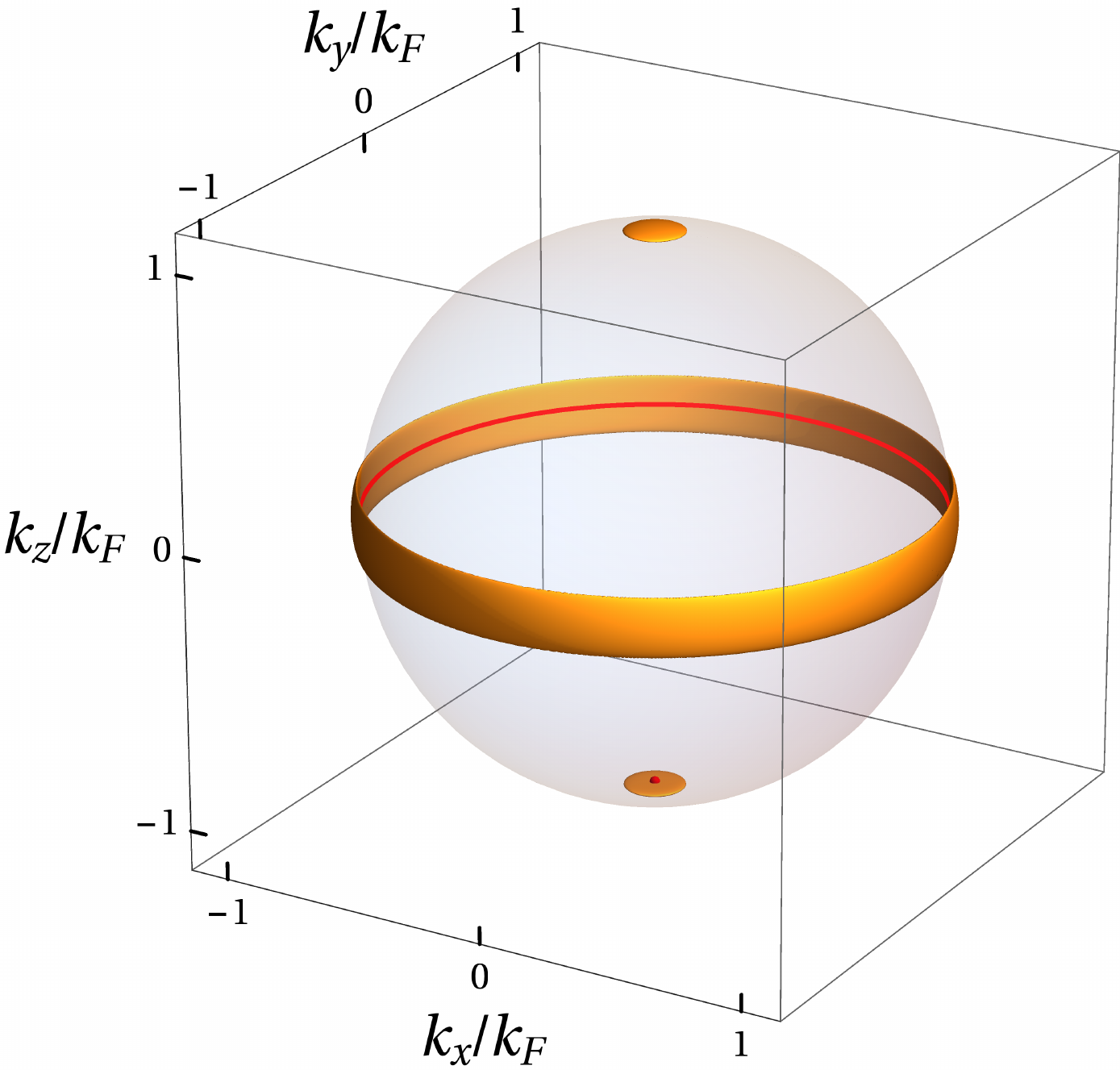}	
\end{center}
\caption{Bogoliubov Fermi surfaces of the superconducting
  $k_z(k_x+ik_y)$ state, shown here for the case where only one band
  has a Fermi surface. The normal-state Fermi surface, shown as the
  semitransparent sphere, is gapped out by the superconductivity. The
  point and line nodes of the single-band theory (red  dots and line,
  respectively), however, are ``inflated'' into spheroidal and toroidal
  $\mathbb{Z}_2$-protected Fermi surfaces (orange surfaces).}
\label{figure1}	
\end{figure}

The superconducting state is taken to have even parity.
Fermionic antisymmetry permits six possible
gap matrices $\eta_i$ in the spin-$3/2$ space:
$\eta_s=U_T$, $\eta_{yz}=(J_y J_z+J_z J_y) U_T/\sqrt{3}$,
$\eta_{xz}=(J_x J_z+J_z J_x)U_T/\sqrt{3}$,
$\eta_{xy}=(J_x J_y+J_y J_x) U_T/\sqrt{3}$,
$\eta_{3z^2-r^2}=(2J_z^2-J_x^2-J_y^2) U_T/3$, and
$\eta_{x^2-y^2}=(J_x^2-J_y^2) U_T/\sqrt{3}$,
where
\begin{equation}
 U_T = \left(
      \begin{array}{cccc}
        0 & 0 & 0 & 1 \\
        0 & 0 & -1 & 0 \\
        0 & 1 & 0 & 0 \\
        -1 & 0 & 0 & 0
      \end{array}
    \right)
\end{equation}%
is the unitary part of the time-reversal operator $T=U_T\mathcal{K}$,
with $\mathcal{K}$ denoting complex conjugation.
The $\eta_s$ gap is a spin-singlet state
and represents pure intraband pairing. The other
gaps, however, describe spin-quintet ($J=2$) pairs and
 involve both intra- and interband pairing~\cite{bry16}.
 Since we consider zero-momentum Cooper pairs, this implies
that quintet pairing involves states away from the Fermi energy.
A general superconducting state is a linear
combination of these gap matrices with symmetry-com\-pli\-ant
$\mathbf{k}$-dependent coefficients and is described by the
Bogoliubov-de Gennes Hamiltonian
\begin{equation}
H(\mathbf{k}) =\left(
      \begin{array}{cc}
        H_N({\bf k}) & \Delta({\bf k}) \\
        \Delta^{\dagger}({\bf k}) & -H_N^T(-\mathbf{k})
      \end{array}
    \right) .
\label{Ham}
\end{equation}
While our results apply to all TRSB even-parity
superconducting states, for concreteness we consider the gap
function
\begin{equation}
\Delta({\bf k})=\Delta_{1}\,\psi({\bf k})\,\eta_s
  +\Delta_0\,(\eta_{xz}+i\eta_{yz}) ,
\label{gap}
\end{equation}
where $\Delta_1$ and $\Delta_0$ are real constants.
Although the latter term describes purely on-site pairing, the gap
matrix $\eta_{xz}+i\eta_{yz}$ transforms under rotations
as the spherical harmonic $Y_{2,1}(\hat{{\bf k}})$,
i.e., it is \emph{chiral}. It
 generically accompanies
a spin-singlet term  with  a form factor
$\psi({\bf k})$ of the same symmetry.
Being chiral, this pairing state, which we call the $k_z(k_x+ ik_y)$
state, breaks time-reversal symmetry. It has the same symmetry as
proposed for URu$_2$Si$_2$~\cite{kas07},
YPtBi~\cite{bry16}, and UPt$_3$ \cite{joy02}.
For pure singlet pairing (i.e., $\Delta_0=0$), the gap has
line nodes in the $k_z=0$ plane and point nodes on the $k_z$ axis
($k_x=k_y=0$). Mixing in a quintet component
 has a dramatic effect on
the excitation spectrum: \emph{the expected point
and line nodes are replaced by Fermi surfaces}. In Fig.~\ref{figure1},
we plot these Fermi surfaces for the $k_z(k_x+ik_y)$ state.
We find them to be a generic feature of all the TRSB even-parity
states classified in \cite{sig91}.
As we will see below, these Fermi surfaces bear some resemblance to
those found in the presence of an exchange field~\cite{gub05},
although they have a completely different origin.

\textit{Existence of Fermi surfaces and $\mathbb{Z}_2$
invariant}.---We now show that Bogoliubov Fermi surfaces are a generic
feature of the Hamiltonian in \eq{Ham} and construct their topological
invariant. The first step is to show that $H(\mathbf{k})$ can be
unitarily transformed into an antisymmetric matrix, i.e., there exists
a unitary $\Omega$ such that
$\tilde{H}^T(\mathbf{k})=-\tilde{H}(\mathbf{k})$ for
$\tilde{H}(\mathbf{k})\equiv \Omega H(\mathbf{k}) \Omega^{\dagger}$.
The main ideas of the proof are explained in the following, details and
a representative $\Omega$ are given in the SM \cite{suppinfo}.
The Hamiltonian $H(\mathbf{k})$ possesses  charge-conjugation
symmetry $C$ and parity symmetry $P$. $C$ acts as
$U_CH^T(-{\bf k})U^{\dagger}_C=-H({\bf k})$ with
$U_C=\hat{\tau}_x \otimes 1_4$, where $\hat{\tau}_i$ are the Pauli
matrices in particle-hole space, while $P$ acts as
$U_PH(-{\bf k})U_P^{\dagger}=H({\bf k})$ with $U_P=\hat{\tau}_0 \otimes
1_4$. Hence, $CP$ symmetry reads
\begin{equation}
  U_{CP}\, H^T(\mathbf{k})\, U_{CP}^\dagger = -H(\mathbf{k}) ,
\label{CPsymmetry}
\end{equation}
with $U_{CP}=U_CU_P^*=\hat{\tau}_x \otimes 1_4$.
This implies that $(CP)^2=U_{CP}U_{CP}^*=+1$ and thus
$U_{CP}=U_{CP}^T$. Any symmetric matrix can be diagonalized by a
unitary congruence, i.e., there exist a unitary $Q$ and a diagonal
$\Lambda$ such that $U_{CP}=Q\Lambda Q^T$ with the \emph{transposed}
matrix $Q^T$. Insertion into \eq{CPsymmetry} gives
$Q \Lambda Q^T\, H^T(\mathbf{k})\, Q^* \Lambda^\dagger Q^\dagger
  = -H(\mathbf{k})$. Since $\Lambda$ is diagonal, we can define a
square root $\sqrt{\Lambda}$. The sign of the root for each diagonal
component of $\Lambda$ can be chosen arbitrarily and is then held
fixed. This allows us to split the unitary matrices in the $CP$ symmetry
relation, which gives
$(\sqrt{\Lambda}{}^\dagger Q^\dagger H(\mathbf{k}) Q \sqrt{\Lambda})^T
  = - \sqrt{\Lambda}{}^\dagger Q^\dagger H(\mathbf{k}) Q \sqrt{\Lambda}$.
This can be written as
$(\Omega H(\mathbf{k}) \Omega^\dagger)^T
  = - \Omega H(\mathbf{k}) \Omega^\dagger$ with
$\Omega \equiv \sqrt{\Lambda}{}^\dagger Q^\dagger$.
Hence, $\tilde{H}(\mathbf{k}) = \Omega H(\mathbf{k})
  \Omega^\dagger$ is antisymmetric, as we wanted to show.
We can then  define the Pfaffian
$P({\bf k})\equiv\text{Pf}\, \tilde{H}({\bf k})$.
Since $\det H({\bf k})=\det \tilde{H}(\mathbf{k})=P^2({\bf k})$,
the zeros of $P({\bf k})$ give the zero-energy states of $H(\mathbf{k})$.

It has recently been shown that Fermi surfaces of Hamiltonians
with $CP$ symmetry squaring to $+1$ can possess a non-trivial
$\mathbb{Z}_2$ charge, making them topologically stable against
$CP$-preserving perturbations \cite{kob14,zha16}. We now express the
invariant in terms of the Pfaffian $P(\mathbf{k})$. The Pfaffian is real
since it is a polynomial of even degree of the components of the matrix
$\tilde{H}(\mathbf{k})$, which is hermitian and antisymmetric and thus
purely imaginary. Regions in momentum
space in which $P(\mathbf{k})$ has opposite  signs are
thus necessarily separated by a two-dimensional (Fermi) surface on which
$P(\mathbf{k})=0$. The existence of such Fermi surfaces is thus guaranteed
if $P(\mathbf{k})$ changes sign and hence we can identify
$(-1)^l=\text{sgn}[P({\bf k}_{-})P({\bf k}_{+})]$ as the
$\mathbb{Z}_2$ invariant, where ${\bf k}_{+}$ (${\bf k}_{-}$) refers to
momenta inside (outside)  of the Fermi surface.

Under what conditions do protected Fermi surfaces exist?
In the normal state, the Pfaffian $P(\mathbf{k}) = \epsilon_+^2
\epsilon_-^2$ is always non-negative and has second-order zeros on
the Fermi surface. Hence, the normal-state Fermi
surface is not protected by the $\mathbb{Z}_2$ charge.
Furthermore, for \emph{time-reversal-symmetric} superconductivity,
$P(\mathbf{k})$ can also be chosen non-negative for all $\mathbf{k}$
so that there is no non-trivial $\mathbb{Z}_2$ invariant
\cite{kob14,zha16}. Our proof in the SM~\cite{suppinfo}
simplifies the earlier proof  by Kobayashi
\textit{et al.}\ \cite{kob14}.
For TRSB pairing, the second-order zeros of $P(\mathbf{k})$ are
generically (i.e., in the absence of additional symmetries or fine tuning)
lifted, leading to $P(\mathbf{k}) > 0$ in a
neighborhood of the former zero, or split into first-order zeros,
in which case there is a region with $P(\mathbf{k}) < 0$.
Such a region is bounded by a two-dimensional Fermi surface; as this
shrinks to a point or line node in the limit of infinitesimal pairing,
we call it an ``inflated'' node.

The existence of protected Bogoliubov Fermi surfaces is now
illustrated for the $k_z(k_x+ik_y)$ state of~\eq{gap}. We restrict
ourselves to pure quintet pairing but our results hold for any
gap with nonzero quintet component,
see the SM \cite{suppinfo}. The Pfaffian is
\begin{equation}
P({\bf k}) = \epsilon_+^2\epsilon_{-}^2 + 4 \Delta_0^2\,
  (\epsilon_+\epsilon_{-}+c_{xz}^2+c_{yz}^2) ,
\label{Pf}
\end{equation}
which is negative for all $\mathbf{k}$ such that
$s_-<\epsilon_{+}\epsilon_{-}<s_+$, where
$s_\pm= -2\Delta_0^2 \pm 2\Delta_0\,(\Delta_0^2-c_{xz}^2-c_{yz}^2)^{1/2}$.
$s_+$ and $s_-$ exist and are distinct if the radicand is positive.
In the plane $k_z=0$ and along the line $k_x=k_y=0$,
this holds for \emph{any} $\Delta_0>0$ since symmetry dictates
that $c_{xz}^2+c_{yz}^2$ vanishes there. We then find $s_-=-4\Delta_0^2<0$
and $s_+=0$, i.e., $s_+$ vanishes on the normal-state Fermi surface.
Since $\epsilon_{+}\epsilon_{-}$ changes sign across the normal-state
Fermi surface, there is always a region with
$s_-<\epsilon_{+}\epsilon_{-}<s_+$ in this plane and along this line and,
due to the continuity of the $c_i(\mathbf{k})$, also in  their
neighborhood. The resulting region with $P(\mathbf{k})<0$ is
bounded by a Fermi surface, as illustrated in Fig.~\ref{figure1}.


\textit{Stability of Fermi surfaces}.---Even-parity TRSB
superconductors are usually argued to be energetically favored over
time-reversal-symmetric states because they maximize the gap in
momentum space \cite{sig91}. The existence of extended
Bogoliubov Fermi surfaces invalidates this
argument. To show that  TRSB states can nevertheless be stable, we
consider a  model with an on-site pairing interaction of
strength $V$ in both the quintet $\eta_{xz}$ and $\eta_{yz}$
channels. In this case, the TRSB state
$\Delta_0\,(\eta_{xz}+i\eta_{yz})$ [i.e., the $k_z(k_x+ik_y)$
state introduced above] and the time-reversal symmetric state
$\sqrt{2}\,\Delta_0\,\eta_{xz}$ have the same critical
temperature $T_c$. To decide which is energetically stable at
temperatures $T$ near $T_c$,
we perform a standard expansion of the free energy $F$ in the gap
$\Delta$~\cite{ho99,min98},
\begin{align}
F &=\frac{1}{2V}\, \text{Tr}\,\Delta^{\dagger}\Delta \notag \\
  & + \frac{k_B T}{2} \sum_{\mathbf{k},\omega_n} \sum_{l=1}^{\infty}
  \frac{1}{l}\, \text{Tr}\, \big[\Delta\tilde{G}(\mathbf{k},\omega_n)
  \Delta^\dagger G(\mathbf{k},\omega_n)\big]^l ,
  \label{eq:freeE}
\end{align}
where $G(\mathbf{k},\omega_n)$ and $\tilde{G}(\mathbf{k},\omega_n)$
are the normal-state electron and hole Matsubara Green's functions,
respectively. For vanishing spin-orbit coupling $\beta$, introduced
below \eq{Ham0}, the normal bands are fourfold degenerate. As was
previously shown in the context of $j=3/2$ pairing in cold
atoms~\cite{ho99}, the TRSB state is \emph{unstable}
towards the time-reversal symmetric state in this limit, as it leaves
two of these bands ungapped. Nonzero spin-orbit coupling partially
lifts this degeneracy and allows the TRSB state to open a gap on all
Fermi surfaces, reducing its energy. An analysis of the fourth-order
term in~\eq{eq:freeE} predicts that the TRSB
$\eta_{xz}+i\eta_{yz}$ state is energetically favored for
$|\beta|k_F^2/k_BT_c \gtrsim 9.324$.
Details are given in the SM~\cite{suppinfo}.
Hence, the presence of the Bogoliubov
Fermi surfaces does not necessarily compromise the stability of
TRSB states.

\textit{Pairing-induced pseudomagnetic field}.---To gain additional
insight into the  Bogoliubov Fermi surfaces, it is useful to rewrite
the Hamiltonian in a basis for which the normal-state Hamiltonian is
diagonal. We denote the eigenvectors of $H_N$ to the twofold
degenerate eigenvalues $\epsilon_\pm$ by $|\pm,i{=}1,2\rangle$,
which we choose to form a pseudospin basis, i.e.,
$|\pm,2\rangle = PT\,|\pm,1\rangle$. In this basis, the
superconducting state is described by the Hamiltonian
\begin{align}
\bar{H}&=\left(\begin{array}{cccc}
    H_{N,+} & \Delta_{++} & 0 & \Delta_{+-} \\
    \Delta_{++}^{\dagger} & -H_{N,+} & -\Delta_{+-}^* & 0 \\
    0 & -\Delta_{+-}^T & H_{N,-} & \Delta_{--} \\
    \Delta_{+-}^{\dagger} & 0 & \Delta_{--}^{\dagger} & -H_{N,-}
  \end{array}\right) .
\label{eq:bandHam}
\end{align}
Here, $H_{N,\pm}=\epsilon_{\pm}\hat{\sigma}_0$ and $\Delta_{\pm\pm}$
are antisymmetric matrices with
$\Delta_{\pm\pm}=\psi_{\pm}({\bf k})\,i\hat{\sigma}_y$,
where $\hat{\sigma}_\mu$ are the Pauli matrices in pseudospin space.
 $\Delta_{+-}$ is the interband pairing potential,
the explicit form of which  depends on the choice of
bases in the two-dimensional eigenspaces of $\epsilon_\pm$ and is
hence not illuminating.
The  intraband gap functions $\psi_{\pm}({\bf k})$ are obtained by
transforming
\eq{gap} into the pseudospin basis and are given by
\begin{equation}
\psi_\pm({\bf k})=\Delta_1\psi({\bf k})\pm2\Delta_0\frac{c_{xz}({\bf
    k})+ic_{yz}({\bf k})}{\epsilon_+({\bf k}) -\epsilon_-({\bf k})} .
\end{equation}
In the absence of $\Delta_{+-}$,
$\bar{H}$ would describe two decoupled pseudospin-$1/2$ singlet
superconductors with, at most, line or point nodes. Hence, the
interband pairing is responsible for the appearance of the
extended Fermi surfaces. This can be shown by treating the off-diagonal
interband blocks of~\eq{eq:bandHam} as a perturbation to the intraband
Hamiltonians: focusing on the $+$ states (analogous results can be found
for the $-$ states  if they have a normal-state Fermi surface),
the second-order corrections due to the interband pairing appear
only in the normal-state components, which become
\begin{equation}
H^\prime_{N,+}=[\epsilon_++\gamma({\bf k})]\, \hat{\sigma}_0
  + {\bf h}({\bf k})\cdot \hat{\mbox{\boldmath$\sigma$}},
\end{equation}
where
\begin{align}
  \gamma({\bf k}) & = \frac{2|\Delta_0|^2}{(\epsilon_+-\epsilon_-)^3}\,
  \left[(\epsilon_+-\epsilon_-)^2-2c_{yz}^2-2c_{xz}^2\right] ,\\
 |{\bf h}({\bf k})| & =\frac{4|\Delta_0|^2}{(\epsilon_+-\epsilon_-)^2}\,
 \sqrt{c_{3z^2-r^2}^2+c_{x^2-y^2}^2+c_{xy}^2} .
 \label{field}
\end{align}
The direction of $\mathbf{h}(\mathbf{k})$ is  basis-dependent
and is thus not
physically meaningful. The correction $\gamma({\bf k})$ is always present
and results in a small modification of the normal-state dispersion
$\epsilon_+$, whereas the second term
${\bf h}({\bf k})\cdot\hat{\mbox{\boldmath$\sigma$}}$
only appears for TRSB gaps. This reveals that in the TRSB state the
\emph{interband pairing manifests itself as a pseudomagnetic field in the
normal-state Hamiltonian}.



The origin of the extended Fermi surfaces becomes clear from
the excitation spectrum of the low-energy pairing Hamiltonian,
\begin{equation}
E_{{\bf k},\pm,\nu}= \nu|{\bf h}({\bf k})|\pm\sqrt{[\epsilon_+({\bf k})
  + \gamma({\bf k})]^2+|\psi_+({\bf k})|^2} ,
\label{Epseudo}
\end{equation}
where $\nu=\pm1$. The pseudomagnetic field $|\mathbf{h}(\mathbf{k})|$
evidently splits the dispersion. The square root in~\eq{Epseudo} goes to
zero at the intersection of the nodes of the intraband gap
$\psi_+({\bf k})$ with the surface $\epsilon_++\gamma({\bf k})=0$,
and the pseudomagnetic field gives rise to the Bogoliubov Fermi surfaces
by shifting the nodes to finite energies $\pm|\mathbf{h}(\mathbf{k})|$.
The generation of ${\bf h}({\bf k})$ by the superconducting state itself
exhibits the \textit{intrinsic} nature of the Bogoliubov Fermi
surfaces. This distinguishes our results from the
``breached-pairing'' state of
population-imbalanced cold atoms or
exchange-split superconductors, where the required breaking of
time-reversal symmetry is \textit{extrinsic} to the
pairing~\cite{liu03,gub05}.

The low-energy effective model also allows us to estimate the dimensions
of the Fermi surfaces. Perpendicular to the normal-state Fermi surface,
the Bogoliubov Fermi surfaces have a width
$\delta k_\perp/k_F \sim \Delta^2_0/[\mu(\epsilon_+-\epsilon_-)]$,
where $k_F$ is the normal-state Fermi momentum. Their width in the
direction parallel to the normal-state Fermi surface is $\delta
k_\parallel/k_F \sim \Delta_0/(\epsilon_+-\epsilon_-)$. Since we
typically expect $\Delta_0\ll \mu$, we find that
$\delta k_\perp \ll \delta k_\parallel$, which implies oblate spheroidal
Fermi surfaces near the original point nodes and flattened toroidal
Fermi surfaces near the original line nodes, as seen in
Fig.~\ref{figure1}.

Pseudomagnetic fields appear in any TRSB phase of our model with
interband (spin-quintet) pairs. Our analysis generalizes to other
 systems with multiband pairing:
we expect non-vanishing contributions
to the pseudomagnetic field from all the interband potentials,
implying that Bogoliubov Fermi surfaces are a generic feature of such
systems.  Equation (\ref{field}) shows that these Fermi surfaces
will be largest in strong-coupling materials in which the different
bands lie close to each other. These conditions are likely satisfied in
heavy-fermion superconductors such as UPt$_3$, Th-doped UBe$_{13}$,
PrOs$_4$Sb$_{12}$, and URu$_2$Si$_2$, which makes them ideal
systems in which to search for Bogoliubov Fermi surfaces.
Indeed, as mentioned in the Introduction, Th-doped UBe$_{13}$
shows a large (and so far unexplained) residual density of
states \cite{zie04} which is consistent with our theory.

\textit{Conclusions}.---We have established that broken-time-reversal
even-parity superconductors generically support two-dimensional Fermi
surfaces. The states at these Fermi surfaces are charge-neutral
Bogoliubov quasiparticles. The Fermi surfaces are protected by a
topological $\mathbb{Z}_2$ invariant, which can be written in terms
of a Pfaffian, and thus cannot be removed by any perturbation that is
$CP$ invariant. They are also energetically stable for plausible
parameters, i.e., the corresponding state has lower free energy than
an associated time-reversal-symmetric state. The
$\mathbb{Z}_2$-protected Fermi surfaces appear in multiband systems,
where interband pairing produces an effective pseudomagnetic field,
which ``inflates'' the expected point or line nodes. Since all
candidate materials for TRSB even-parity superconductivity have
multiple relevant bands, we expect that these
$\mathbb{Z}_2$-protected Fermi surfaces are ubiquitous. The existence
of Bogoliubov Fermi surfaces in the superconducting state should lead
to characteristic experimental consequences: a nonzero density of
states at the Fermi energy, visible for example in tunneling and
photoemission experiments, would coexist with ideal conductivity and
flux expulsion. The low-temperature thermodynamic response would show
a linear temperature dependence of the specific heat, and heat
conduction should also be unconventional~\cite{gub05}. Such anomalies
may have already been observed in heavy-fermion superconductors
\cite{zie04,joy02}. Our work raises many interesting new questions,
e.g., is a superconductor with Bogoliubov Fermi surfaces a Fermi
liquid when residual interactions are taken into account? The search
for and study of such systems thus constitutes a very promising task
for future research.

The authors thank A.\,P. Schnyder and M. Punk for useful discussions.
D.\,F.\,A. was supported by the National Science Foundation grant
DMR-1335215. C.\,T. acknowledges financial support by the Deut\-sche
For\-schungs\-ge\-mein\-schaft, in part through Research Training
Group GRK 1621 and Collaborative Research Center SFB 1143.

\onecolumngrid


\clearpage
\onecolumngrid

\setcounter{equation}{0}
\setcounter{page}{1}

\begin{center}
\textbf{\large  Supplemental Material for\\[1ex]
Bogoliubov Fermi Surfaces in Superconductors with
Broken Time-Reversal Symmetry}\\[2ex]
 D. F. Agterberg, P. M. R. Brydon, and C. Timm
\end{center}

\section{Mapping to a general two-orbital model}

In this section, we show that the most general model for a
normal-state system with a two-valued orbital degree of freedom where
the orbitals have the same parity and are chosen to be real
can be unitarily transformed into the $j=3/2$ Hamiltonian given in
the main text. We expand the single-particle Hamiltonian in terms of
Kronecker products in orbital-spin space,
\begin{equation}
H_N({\bf k}) = \sum_{\mu,\nu=0,x,y,z} c_{\mu,\nu}({\bf k})\,
  \hat{s}_{\mu} \otimes \hat{\sigma}_{\nu} ,
\label{eq:suppH}
\end{equation}
where $\hat{s}_{\mu}$ ($\hat{\sigma}_{\nu}$) are the Pauli matrices
in orbital (spin) space. Hermiticity implies
$c_{\mu,\nu}({\bf k})=c_{\mu,\nu}^*({\bf k})$. Inversion maps
$H_N(\mathbf{k})$ onto $U_P H_N(-\mathbf{k}) U_P^\dagger$, where
$U_P = 1_4$ ($U_P = -1_4$) if the orbitals are both even (odd). The
extra sign for the odd case obviously drops out and can be disregarded.
Hence, for the Hamiltonian to be symmetric under inversion, we simply
require $c_{\mu,\nu}({\bf k})=c_{\mu,\nu}(-{\bf k})$. Time reversal
maps $H_N(\mathbf{k})$ onto $U_T H_N^*(-\mathbf{k}) U_T^\dagger$.
$c_{\mu,\nu}({\bf k})$ is thus mapped onto
$c_{\mu,\nu}^*(-{\bf k})=c_{\mu,\nu}({\bf k})$. Under the assumption of
real-valued orbitals, $\hat{s}_y$ and the spin components
$\hat{\sigma}_x$, $\hat{\sigma}_y$, $\hat{\sigma}_z$ are odd under time
reversal, whereas the remaining matrices are even. This is achieved by
$U_T = \hat{s}_0 \otimes i\hat{\sigma}_y$.

In order for the Hamiltonian $H_N(\mathbf{k})$ to be invariant under time
reversal, only those Kronecker products can appear in \eq{eq:suppH} that
are even, specifically the six products with
$\{\mu,\nu\} = \{0,0\}, \{x,0\}, \{z,0\}, \{y,x\}, \{y,y\}, \{y,z\}$. The
corresponding $c_{\mu,\nu}({\bf k})$ are real and even functions of
${\bf k}$. Note that the five nontrivial Kronecker products are mutually
anticommuting and thus form a representation of the five Dirac matrices.

We now show that the generalized Luttinger-Kohn model \cite{S:lut55}
adopted in the main text is equivalent to~\eq{eq:suppH}. It is expressed
in terms of the  angular-momentum $j=3/2$ matrices
\begin{align}
J_x &= \frac{1}{2}\left(
      \begin{array}{cccc}
        0 & \sqrt{3} & 0 & 0 \\
        \sqrt{3} & 0 & 2 & 0 \\
        0 & 2 & 0 & \sqrt{3} \\
        0 & 0 & \sqrt{3} & 0
      \end{array}
    \right) ,\\
J_y &= \frac{i}{2}\left(
      \begin{array}{cccc}
        0 & -\sqrt{3} & 0 & 0 \\
        \sqrt{3} & 0 & -2 & 0 \\
        0 & 2 & 0 & -\sqrt{3} \\
        0 & 0 & \sqrt{3} & 0
      \end{array}
    \right) ,\\
J_z &= \frac{1}{2}\left(
      \begin{array}{cccc}
        3 & 0 & 0 & 0 \\
        0 & 1 & 0 & 0 \\
        0 & 0 & -1 & 0 \\
        0 & 0 & 0 & -3
      \end{array}
    \right) .
\end{align}
The generalized Luttinger-Kohn model involves five sets of products of the
$j=3/2$ matrices, specifically  $(2 J_z^2 - J_x^2 - J_y^2/6$,
$(J_x^2 - J_y^2)/(2\sqrt{3})$, $(J_x J_z + J_z J_x)/(2\sqrt{3})$,
$(J_x J_y + J_y J_x)/(2\sqrt{3})$, and $(J_y J_z + J_z J_y)/(2\sqrt{3})$.
These matrices also form a set of five Dirac matrices and there is a
unitary transformation between them and the five allowed nontrivial
Kronecker products discussed above.  Specifically, we find
\begin{align}
 U^\dagger\, \frac{2 J_z^2 - J_x^2 - J_y^2}{6}\, U
  &= \hat{s}_z\otimes \hat{\sigma}_0 , \\
 U^\dagger\, \frac{J_x^2 - J_y^2}{2\sqrt{3}}\, U
  &= \hat{s}_x\otimes \hat{\sigma}_0 , \\
U^\dagger\, \frac{J_x J_z + J_z J_x}{2\sqrt{3}}\, U
  &= \hat{s}_y\otimes \hat{\sigma}_y , \\
U^\dagger\, \frac{J_x J_y + J_y J_x}{2\sqrt{3}}\, U
  &= \hat{s}_y\otimes \hat{\sigma}_z , \\
U^\dagger\, \frac{J_y J_z + J_z J_y}{2\sqrt{3}}\, U
  &= \hat{s}_y\otimes \hat{\sigma}_x ,
\end{align}
with the unitary matrix
\begin{equation}
 U=\left(
      \begin{array}{cccc}
        1 & 0 & 0 & 0 \\
        0 & 0 & 0 & 1 \\
        0 & 0 & 1 & 0 \\
        0 & 1 & 0 & 0
      \end{array}
    \right) .
\end{equation}
Note that one can pairwise swap the orbital-spin matrices
$\hat{s}_\mu\otimes\hat{\sigma}_\nu$ by an appropriate unitary
transformation so that this mapping is entirely general. We hence see that
the generalized Luttinger-Kohn model is equivalent to the most general
single-particle Hamiltonian of a time- and inversion-symmetric material
with two identical-parity orbitals and spin. The generalized
Luttinger-Kohn representation is convenient, however, as it allows us to
exploit the symmetry properties of the  angular-momentum matrices $J_i$
under rotations to gain insight into the underlying physics.

We can obtain six on-site pairing states in the orbital-spin
representation by multiplying the six allowed Kronecker products by the
unitary part of the time-reversal operator,
$U_T=\hat{s}_0\otimes i\hat{\sigma}_y$. We hence obtain one
orbitally trivial spin-singlet
gap proportional to $\hat{s}_0\otimes \hat{\sigma}_y$ and five
``anomalous'' gaps proportional to $\hat{s}_x\otimes \hat{\sigma}_y$,
$\hat{s}_z\otimes \hat{\sigma}_y$, $\hat{s}_y\otimes\hat{\sigma}_z$,
$\hat{s}_y\otimes \hat{\sigma}_0$, and $\hat{s}_y\otimes
\hat{\sigma}_x$, which are either orbital-triplet spin-singlet or
orbital-singlet spin-triplet pairing states.
It is straightforward to map these states onto the gap matrices in the
equivalent spin-$3/2$ formulation: the orbitally trivial spin-singlet
state maps onto the singlet gap matrix
$\eta_s$, while the anomalous gaps map onto the quintet gap matrices.
Hence, we can describe
any even-parity pairing state, without loss of generality, in terms of
a linear combination of the six spin-$3/2$ gap functions with even-parity
coefficients $\psi_i({\bf k})$.

\section{Existence and properties of the Pfaffian}

In this section,  we provide additional details on
constructing the topological invariant protecting the Fermi surfaces.
Our starting point is the Hamiltonian in the superconducting state,
\begin{equation}
H(\mathbf{k}) = \left(
      \begin{array}{cc}
        H_N({\bf k}) & \Delta({\bf k}) \\
        \Delta^{\dagger}({\bf k}) & -H_N^T({\bf k})
      \end{array}
    \right) .
\label{eq:Ham3}
\end{equation}
Recall that $H_N(\mathbf{k})$ is even in $\mathbf{k}$. We employ the
spin-$3/2$ basis used in the main text. We first  show that a
$\mathbf{k}$-independent unitary matrix $\Omega$ exists such that
$\tilde{H}^T(\mathbf{k})=-\tilde{H}(\mathbf{k})$ for
$\tilde{H}({\bf k})=\Omega H({\bf k}) \Omega^{\dagger}$, i.e., the
Hamiltonian can be transformed into antisymmetric form.

The proof proceeds as follows: 1.~The Hamiltonian $H(\mathbf{k})$
satisfies parity and charge-conjugation symmetries and thus also their
product, i.e.,
\begin{equation}
U_{CP}\, H^T(\mathbf{k})\, U_{CP}^\dagger = -H(\mathbf{k}) ,
\label{1.sup.1}
\end{equation}
where $U_{CP} \equiv U_C U_P^* = (\hat{\tau}_x \otimes 1_4)
  (\hat{\tau}_0 \otimes 1_4) = \hat{\tau}_x \otimes 1_4$.
The $\hat{\tau}_i$ denote Pauli matrices in particle-hole
(Nambu) space. We find that the $CP$ symmetry squares to $+1$ since
$(CP)^2 = (U_{CP} \mathcal{K})^2
  = U_{CP} U_{CP}^* = \hat{\tau}_0 \otimes 1_4 = +1_8$.
In the presence of such a symmetry, two-dimensional Fermi surfaces are
characterized by a $\mathbb{Z}_2$ invariant~\cite{S:kob14,S:zha16}.

2.~For \emph{any} $CP$ symmetry that squares to unity we have
$U_{CP}^* = U^{-1}_{CP} = U^\dagger_{CP}$ and thus
$U_{CP} = U^T_{CP}$, hence $U_{CP}$ is symmetric. Any (complex)
symmetric matrix can be diagonalized by a unitary \emph{congruence},
i.e., there  exist a unitary matrix $Q$ and a diagonal matrix
$\Lambda$ such that (note the transpose)
\begin{equation}
U_{CP} = Q \Lambda Q^T .
\label{1.UCP.5}
\end{equation}
Inserting this equation into \eq{1.sup.1} yields
\begin{equation}
Q \Lambda Q^T\, H^T(\mathbf{k})\, Q^* \Lambda^\dagger Q^\dagger
  = -H(\mathbf{k}) .
\label{1.CPQ.2}
\end{equation}
Since $\Lambda = Q^\dagger U_{CP} Q^*$ is unitary and diagonal, it can be
written as $\Lambda = \mathrm{diag}(\lambda_1,\lambda_2,\ldots)$ with
$|\lambda_i|=1$. Now let
\begin{equation}
\sqrt{\Lambda} \equiv
  \mathrm{diag}\left(\sqrt{\lambda_1},\sqrt{\lambda_2},\ldots\right) ,
\label{1.Lambda.4}
\end{equation}
where for each $i$, $\sqrt{\lambda_i}$ is the complex root with arbitrary but
fixed sign. Also let
\begin{equation}
\sqrt{\Lambda}{}^\dagger \equiv \mathrm{diag}
  \left(\sqrt{\lambda_1}^{\,*},\sqrt{\lambda_2}^{\,*},\ldots\right) ,
\label{1.Lambda.5}
\end{equation}
with the same choice of signs as in \eq{1.Lambda.4}. It is trivial to show
that this is a root of $\Lambda^\dagger$. Furthermore, $\sqrt{\Lambda}$ and
 $\sqrt{\Lambda}{}^\dagger$ are obviously diagonal, and thus symmetric,
and also unitary. We can thus rewrite \eq{1.CPQ.2} as
$Q \sqrt{\Lambda} \sqrt{\Lambda} Q^T H^T(\mathbf{k})
  Q^* \sqrt{\Lambda}{}^\dagger \sqrt{\Lambda}{}^\dagger Q^\dagger
  = -H(\mathbf{k})$
and find that
\begin{equation}
(\sqrt{\Lambda}{}^\dagger Q^\dagger H(\mathbf{k}) Q \sqrt{\Lambda})^T
  = - \sqrt{\Lambda}{}^\dagger Q^\dagger H(\mathbf{k}) Q \sqrt{\Lambda} .
\label{1.CPQ.4}
\end{equation}
With the unitary matrix
$\Omega \equiv \sqrt{\Lambda}{}^\dagger Q^\dagger$, \eq{1.CPQ.4}
can be written as
$(\Omega H(\mathbf{k}) \Omega^\dagger)^T
  = - \Omega H(\mathbf{k}) \Omega^\dagger$.
Hence, $\tilde{H}(\mathbf{k}) \equiv \Omega H(\mathbf{k})
  \Omega^\dagger$ is indeed antisymmetric.

For the Hamiltonian $H(\mathbf{k})$ given in \eq{eq:Ham3}
above together with Eq.\ ({1}) in the main text, a possible choice is
\begin{equation}
\Omega = \frac{1}{\sqrt{2}} \left(\begin{array}{cc}
  1 & 1 \\ i & -i
  \end{array}\right) \otimes 1_4 .
\end{equation}
This yields
\begin{equation}
\tilde{H}(\mathbf{k}) = \frac{1}{2} \left(\begin{array}{cc}
  H_N - H_N^T + \Delta + \Delta^\dagger &
    -i\,(H_N+H_N^T) + i\,(\Delta-\Delta^\dagger) \\[1ex]
  i\,(H_N+H_N^T) + i\,(\Delta-\Delta^\dagger) &
    H_N - H_N^T - \Delta - \Delta^\dagger \end{array}\right) ,
\end{equation}
where we have suppressed the argument $\mathbf{k}$. The terms involving
the normal-state Hamiltonian $H_N$ are obviously antisymmetric. For
this, it is crucial that $H_N$ is even in $\mathbf{k}$. From \eq{1.sup.1},
one easily finds $\Delta^T = -\Delta$, which implies that also the
superconducting terms in $\tilde{H}(\mathbf{k})$ are antisymmetric. Hence,
we do find $\tilde{H}^T(\mathbf{k}) = -\tilde{H}(\mathbf{k})$.

Since $\tilde{H}(\mathbf{k})$ is antisymmetric, its Pfaffian
$P({\bf k}) \equiv \text{Pf}\, \tilde{H}({\bf k})$ exists.
The Pfaffian is real for any spinful system since (a) the dimension of the
Hamiltonian is a multiple of four ($2\times2$ for Nambu and spin space), thus
the Pfaffian is a polynomial of even degree of the components of
$\tilde{H}(\mathbf{k})$, and (b) $\tilde{H}(\mathbf{k})$ is hermitian and
antisymmetric and thus these components are purely imaginary.

Note further that the ambiguity in the signs of the roots in $\sqrt{\Lambda}$
means that $\Omega$ can be multiplied on the left by  a diagonal matrix $D$ with
arbitrary components $\pm 1$ on the diagonal. With $\Omega \to D\Omega$ we get
$\Omega^\dagger \to \Omega^\dagger D$ and $\tilde{H}(\mathbf{k}) \to D
\tilde{H}(\mathbf{k}) D$. This leads to $\text{Pf}\,\tilde{H}(\mathbf{k}) \to
  \det D\: \text{Pf}\,\tilde{H}(\mathbf{k})$.
Hence, $P(\mathbf{k})$ is only determined up to an overall sign. But this sign is
selected by fixing the root $\sqrt{\Lambda}$ once for all $\mathbf{k}$. Thus
sign \emph{changes} in $P(\mathbf{k})$ are meaningful.

Since $\det H(\mathbf{k})=\det \tilde{H}(\mathbf{k})=P^2(\mathbf{k})$,
the zeros of $P(\mathbf{k})$ give the nodes of the superconducting state. Thus if
the Pfaffian changes sign as a function of $\mathbf{k}$, the surface separating
regions with $P(\mathbf{k})\gtrless 0$ is a two-dimensional Fermi surface.
However, $P(\mathbf{k})$ can in addition have zeros of even multiplicity, which
do not separate regions with different sign. These \emph{accidental} zeros can
form two-dimensional surfaces or line or point nodes.

For the general $k_z(k_x+ik_y)$ state of Eq.\ ({4}) in the main text,
the Pfaffian is
\begin{equation}
P({\bf k}) =  (\epsilon_+\epsilon_-)^2
  + 4 \Delta_0^2\, (\epsilon_+\epsilon_-+c_{xz}^2+c_{yz}^2)
  + \Delta_1^2\, |\psi|^2\, (\epsilon_+^2+\epsilon_-^2)
  + 8\Delta_0\Delta_1\, c_0\, |\psi|^2
  + \Delta_1^4\, |\psi|^4  ,
\label{Pf}
\end{equation}
where $\epsilon_\pm(\mathbf{k})$ are the normal-state eigenenergies.
We have chosen the signs in the root $\sqrt{\Lambda}$ in such a
way that $P(\mathbf{k})>0$ in the limit of large $\mathbf{k}$.
Equation (\ref{Pf}) is correct for arbitrary real and even
coefficients $c_i(\mathbf{k})$. Note that when $\Delta_0=0$, the
equation $P(\mathbf{k})=0$ implies
the usual relationship for zero-energy states in a spin-singlet
superconductor, i.e.,
$(\epsilon_+^2+\Delta_1^2\,|\psi|^2)(\epsilon_-^2+\Delta_1^2\,|\psi|^2)=0$,
implying nodes when $\epsilon_{\pm}=0$ and $\psi=0$; since
$\psi(\mathbf{k})\propto c_{xz}(\mathbf{k})+ic_{yz}(\mathbf{k})$ to
ensure that the full gap function $\Delta(\mathbf{k})$ transforms
under rotations like $Y_{2,1}(\hat{\mathbf{k}})$, we have line nodes
in the $k_z=0$ plane and point nodes along the line $k_x=k_y=0$.
With both
$\Delta_0\ne 0$ and $\Delta_1\ne 0$, we see that along the nodal directions,
for which $\psi \propto c_{xz}+ic_{yz}=0$, we have zero-energy excitations
for $\epsilon_+\epsilon_- (4\Delta_0^2+\epsilon_+\epsilon_-)=0$. The two
solutions for $\epsilon_+\epsilon_-$ that follow from the latter equation
give the two points of the Bogoliubov Fermi surface along these nodal
directions. Away from the nodal directions, the other terms in \eq{Pf} no
longer vanish. However, they are continuous functions of the direction
$\hat{\mathbf{k}}$, implying that the Fermi surface still exists at least
for a finite range away from the nodal directions. In the main text, we
consider the special case of a pure quintet gap, $\Delta_1=0$.

Finally, we show that the Pfaffian can be chosen non-negative if the system
is also time-reversal symmetric and the combined inversion and time-reversal
symmetry squares to $-1$, i.e., if there exists a unitary matrix $U_{PT}$ so
that $U_{PT} H^T(\mathbf{k}) U_{PT}^\dagger = H(\mathbf{k})$
and $U_{PT} U^*_{PT} = -1$. Hence, if both symmetries are
present,  the $\mathbb{Z}_2$ invariant exists but is necessarily
trivial. This has  essentially been shown using the method of Clifford
algebra extensions in \cite{S:kob14}. In the following, we give a more
elementary proof.

1.~Kramers' theorem shows that under the last
two conditions all eigenvalues of $H(\mathbf{k})$ have even degeneracy.
Furthermore, condition (\ref{1.sup.1}) above implies that
the eigenvalues come in pairs $\{E_i(\mathbf{k}), -E_i(\mathbf{k})\}$. Since the
dimension of $H(\mathbf{k})$ is a multiple of four, the spectrum thus consists of
quadruplets $\{E_i(\mathbf{k}), E_i(\mathbf{k}), -E_i(\mathbf{k}),
-E_i(\mathbf{k})\}$.

2.~$P(\mathbf{k})$ is a polynomial of the components of $\tilde{H}(\mathbf{k})$,
which are linear combinations of the components of $H(\mathbf{k})$. Hence,
$P(\mathbf{k})$ is a polynomial of the components of $\mathcal{H}(\mathbf{k})$.
The coefficients are independent of $\mathbf{k}$.

3.~We have $P^2(\mathbf{k}) = \det \tilde{H}(\mathbf{k})
  = \det H(\mathbf{k}) = \prod_i E_i^4(\mathbf{k})$ so that
$P(\mathbf{k}) = \pi(\mathbf{k}) \prod_i E_i^2(\mathbf{k})$
with $\pi(\mathbf{k}) = \pm 1$.

4.~Consider the eigenenergies $E_i(\mathbf{k})$ as functions of a real parameter
$h$, which can be any real or imaginary part of a component of the hermitian,
finite-dimensional matrix $H(\mathbf{k})$. Theorem 7.6 of Alekseevsky
\textit{et al}.\ \cite{S:AKM98} shows under the weak additional condition that no
two eigenvalues meet of infinite order for any real $h$ unless they are
equal for all $h$ that the $E_i(\mathbf{k})$ can be chosen as smooth
functions of $h$. Then $\prod_i E_i^2(\mathbf{k})$ is a smooth function of $h$.

5.~We have already shown that $P(\mathbf{k})$ is a polynomial in $h$,
that $\prod_i E_i^2(\mathbf{k}) = \pi(\mathbf{k})\, P(\mathbf{k})$ is a
smooth function of $h$, and that $\pi(\mathbf{k})=\pm 1$. Then
$\pi(\mathbf{k})$ is constant.

6.~Noting the ambiguity in choosing $\Omega$ and thus the overall sign of the
Pfaffian, we can choose $\pi(\mathbf{k})=+1$ at some
$\mathbf{k}$ and consequently at all $\mathbf{k}$. Thus we indeed obtain
$P(\mathbf{k}) = \prod_i E_i^2(\mathbf{k}) \ge 0$.

\section{Stability of states with broken time-reversal symmetry}

 In this section, we provide additional details  on the free-energy
expansion. We show that states with broken time-reversal symmetry featuring
Bogoliubov Fermi surfaces can be energetically stable.
The relative stability of spin-singlet broken-time-reversal states over
time-reversal-invariant states  with the same transition temperature
has been attributed to the gapping of nodes that appear in the
time-reversal-symmetric state by breaking this symmetry \cite{S:sig91}. A
common example is the chiral \textit{d}-wave state with line nodes,
for which the usual spin-singlet gap function takes the form
$\psi({\bf k})=\psi_0\, k_z(\nu_x k_x+\nu_y k_y)$. In the
broken-time-reversal state, we have $(\nu_x,\nu_y)=(1,i)/\sqrt{2}$, which
leads to line nodes for $k_z=0$ and point nodes at $k_x=k_y=0$. In the nodal
time-reversal-invariant state, we instead have $(\nu_x,\nu_y)=(1,0)$, which
also leads to line nodes for $k_z=0$ and additional line nodes for $k_x=0$.
In this case, the broken-time-reversal state is believed to be stable
because it gaps the $k_x=0$ line node so that only two point nodes remain,
gaining condensation energy. However, we have found that these point nodes
become inflated $\mathbb{Z}_2$-protected Fermi surfaces, and it is
reasonable to ask if the broken-time-reversal state is still  stable.


 We consider  the spherically symmetric normal-state Hamiltonian
$H_N(\mathbf{k})=\alpha k^2+\beta\,({\bf k}\cdot {\bf J})^2-\mu$ and an on-site
gap function that is a linear combination of $\eta_{xz}$ and $\eta_{yz}$.
For this gap function, it is known that there are two essentially
different possible ground states \cite{S:sig91}: a
time-reversal-invariant state $\Delta_r=(\bar{\Delta}/2)\,\eta_{xz}$ with
$k_zk_x$ symmetry and a broken-time-reversal state
$\Delta_a=(\bar{\Delta}/2\sqrt{2})\,(\eta_{xz}+i\eta_{yz})$
with $k_z(k_x+ik_y)$ symmetry.
 The time-reversal-invariant state $(\bar{\Delta}/2)\,\eta_{yz}$
is degenerate with $\Delta_r$. For these two states, the off-diagonal
block in the Hamiltonian $H(\mathbf{k})$ reads
\begin{equation}
\Delta_{r}=\frac{\bar{\Delta}}{2}\left(
      \begin{array}{cccc}
        0 & 0 & -1 & 0 \\
        0 & 0 & 0 & 1 \\
        1 & 0 & 0 & 0 \\
        0 & -1 & 0 & 0
      \end{array}\right)
\end{equation}
and
\begin{equation}
\Delta_{a}=\frac{\bar{\Delta}}{\sqrt{2}}\left(
      \begin{array}{cccc}
        0 & 0 & -1 & 0 \\
        0 & 0 & 0 & 0 \\
        \,1\, & 0 & 0 & 0 \\
        0 & 0 & 0 & 0
      \end{array}\right),
\end{equation}
respectively. These two matrices have been normalized such that
$\text{Tr}\,\Delta^\dagger \Delta=\bar{\Delta}^2$ ($\bar{\Delta}$
is assumed real).

Near $T_c$, the mean-field free energy can be expressed as a power series in
 $\Delta$. The expression  is standard \cite{S:ho99,S:min98},
\begin{equation}
F =\frac{1}{2V}\, \text{Tr}\,\Delta^{\dagger}\Delta
  + \frac{k_B T}{2} \sum_{\mathbf{k},\omega_n} \sum_{l=1}^{\infty}
  \frac{1}{l}\, \text{Tr}\, \big[\Delta\tilde{G}(\mathbf{k},\omega_n)
  \Delta^\dagger G(\mathbf{k},\omega_n)\big]^l ,
\end{equation}
where the constant $V$ is the BCS pairing interaction (which is the
same for $\Delta_a$ and $\Delta_r$), $\omega_n$ are the Matsubara
frequencies, $T$ is the temperature, and the normal-state Green's
functions $G$ and $\tilde{G}$ satisfy
 $[i\omega_n-H_N(\mathbf{k})]\, G(\mathbf{k},\omega_n)=1$ and
$[i\omega_n+H_N^T(\mathbf{k})]\, \tilde{G}(\mathbf{k},\omega_n)=1$.
To find $G$ and $\tilde{G}$, we note that
$H_N^T(k_x,k_y,k_z)=H_N(k_x,-k_y,k_z)$ so that $\tilde{G}$ is given
once we know $G$. Some algebra then gives
\begin{align}
G(\mathbf{k},\omega_n)
  &= G_+(\mathbf{k},\omega_n)
  + \left[\big(\hat{\mathbf{k}}\cdot {\bf J}\big)^2-\frac{5}{4}\right]
    G_-(\mathbf{k},\omega_n) , \\
\tilde{G}(\mathbf{k},\omega_n)
  & = \tilde{G}_+(\mathbf{k},\omega_n)
  + \left[\big(\hat{\mathbf{k}}\cdot {\bf J}^T\big)^2-\frac{5}{4}\right]
    \tilde{G}_-(\mathbf{k},\omega_n) ,
\end{align}
with
\begin{align}
G_{\pm}(\mathbf{k},\omega_n)
  &= \frac{1}{2}\left(\frac{1}{i\omega_n-\epsilon_1}
  \pm\frac{1}{i\omega_n-\epsilon_2}\right), \\
\tilde{G}_{\pm}(\mathbf{k},\omega_n)
  & = \frac{1}{2}\left(\frac{1}{i\omega_n+\epsilon_1}
  \pm\frac{1}{i\omega_n+\epsilon_2}\right),
\end{align}
$\epsilon_1=(\alpha+9\beta/4)\,k^2-\mu$, and
$\epsilon_2=(\alpha+\beta/4)\,k^2-\mu$. Denoting the free energies for
$\Delta_r$ and $\Delta_a$ by $F_r$ and $F_a$, respectively, we find, to
fourth order in $\Delta$,
\begin{equation}
F_{a}-F_r \cong \frac{\bar{\Delta}^4k_B T}{16} \sum_{{\bf k},\omega_n}
  \Big\{\tilde{G}_+^2G_+^2
    + (1-2l_1l_{-1})\left(\tilde{G}_-^2G_+^2+\tilde{G}_+^2G_-^2\right)
    + 4(l_1l_{-1}-1)\,\tilde{G}_-G_-\tilde{G}_+G_+
    + (1-l_1^2l_{-1}^2)\, \tilde{G}_-^2G_-^2\Big \} ,
\label{free}
\end{equation}
where $l_{\pm 1}=\sqrt{3}\,\cos\theta\sin\theta\: e^{\pm  i \phi}$ and
$\theta$ and $\phi$ are the spherical angles denoting the direction of
${\bf k}$. Note that the term of order $\Delta^2$ drops out  of the
free-energy difference under the integral over $\phi$.  The analysis
of the above expression reveals that either $\Delta_r$ or $\Delta_a$ can
be stable, depending on the parameters. In particular, first consider the
case of vanishing spin-orbit coupling, $\beta=0$. In this limit, we find
$\tilde{G}_-=G_-=0$ so that only the first term in the sum in \eq{free}
survives, and we have $F_{a}>F_{r}$ so that $\Delta_r$ has lower free
energy. This limit has also been considered in the context of $j=3/2$
pairing in cold atoms \cite{S:ho99,S:mer74} and the results agree with what
we find. Consequently, the broken-time-reversal state is not stable for
$\beta=0$.

  Now consider the single-band limit,  which can be reached, for example,
by taking a large $|\beta|$, such that $\alpha+9\beta/4$ and
$\alpha+\beta/4$ have opposite sign, and a large chemical
potential. If only $\epsilon_1$ crosses the Fermi surface, for
sufficiently small $k_BT/\mu\ll 1$ we can safely take the limit
$|\epsilon_2|/k_BT\rightarrow \infty$ to find the asymptotic
expression
\begin{equation}
F_a - F_r
  \cong \frac{\bar{\Delta}^4}{1024 k_BT} \sum_{{\bf k}} l_1^2l_{-1}^2
  \left( \frac{1-k_BT\sinh(\epsilon_1/k_BT)/\epsilon_1}{\epsilon_1^2(1+\cosh(\epsilon_1/k_BT))}
  \right) < 0\,.
\end{equation}
This is the result expected from single-band weak-coupling
theory, and demonstrates that the broken time-reversal state is stable
in this limit.

The above analysis implies that for fixed $\alpha$ and $\mu$ there will be
a transition from $\Delta_a$ to $\Delta_r$ as a function of spin-orbit
coupling $\beta$. It is straightforward to show that the momentum and
Matsubara sum in \eq{free} only depends on the dimensionless ratios
$\beta/\alpha$ and $\mu/k_BT$, apart from prefactors that do not affect
its sign. Numerical evaluation indicates that transitions exist for any
value of $\mu/k_BT$. We consider explicitly the case of
$|\beta|\ll \alpha$, in which the two spherical Fermi surfaces have nearly
identical $k_F$. We then find by expanding in $\beta/\alpha$ that this
transition takes place at
$x \equiv |\beta|\mu/\alpha k_BT = |\beta|k_F^2/k_BT \approx 9.324$. For
$x \lesssim 9.324$, we find that the time-reversal-invariant state
$\Delta_r$ is the ground state, while for $x \gtrsim 9.324$, $\Delta_r$ is
unstable towards $\Delta_a$. This indicates that  the
broken-time-reversal state is stabilized by modest spin-orbit coupling.

\end{document}